\documentclass[%
 reprint,
 superscriptaddress,
 amsmath,amssymb,
 aps,
]{revtex4-1}

\usepackage{graphicx}
\usepackage{subfigure}
\renewcommand{\thesubfigure}{\alph{subfigure}}
\makeatletter
\renewcommand{\@thesubfigure}{(\thesubfigure)\space}
\makeatother
\usepackage{dcolumn}
\usepackage{bm}
\usepackage[colorlinks, linkcolor=blue, citecolor=blue, urlcolor=blue]{hyperref}

\usepackage[normalem]{ulem}

\begin{document}


\title{Measurement-induced phase transition: 
A case study in the non-integrable model by density-matrix renormalization group calculations}

\author{Qicheng Tang}
\email{tangqicheng@westlake.edu.cn}

\affiliation{%
	Zhejiang University, Hangzhou 310027, China
}%
\affiliation{%
	School of Science, Westlake University, Hangzhou 310024, China
}%
\affiliation{%
	Institute of Natural Sciences, Westlake Institute of Advanced Study, Hangzhou 310024, China
}%

\author{W. Zhu}
\email{zhuwei@westlake.edu.cn}
 
\affiliation{%
	School of Science, Westlake University, Hangzhou 310024, China
}%
\affiliation{%
	Institute of Natural Sciences, Westlake Institute of Advanced Study, Hangzhou 310024, China
}%

\date{\today}

\begin{abstract}
	We study the effect of local projective measurements on the quantum quench dynamics.
	As a concrete example, a one-dimensional Bose-Hubbard model is simulated by the matrix product state and time-evolving block decimation.
	We map out a global phase diagram in terms of the measurement rate in spatial space and time domain,
	which demonstrates a volume-to-area law entanglement phase transition.
	When the measurement rate reaches the critical value,
	we observe a logarithmic growth of entanglement entropy as the subsystem size or evolved time increases.
	Moreover, we find that the probability distribution of the single-site entanglement entropy distinguishes the volume and area law phases, similar to the case of disorder-induced many-body localization.
	We also investigate the scaling behavior of entanglement entropy and mutual information between two separated sites, 
	which is indicative of a single universality class and thus suggests a possible unified description of this transition.


\end{abstract}

\pacs{Valid PACS appear here}
\maketitle


\section{Introduction}

Quantum entanglement is an invaluable tool to access the intrinsic nature of underlying states
and their non-equilibrium properties in quantum physics \cite{Amico2008, Eisert2010, laflorencie2016quantum}.
For example,  typical excited eigenstates exhibit a volume-law scaling; i.e.,
the entanglement entropy of the reduced state of a sub-system grows as its volume. 
In contrast, for most gapped ground states, an area-law of entanglement entropy emerges 
(entropy is proportional to the surface area of the sub-system).
The scaling behavior of entanglement entropy could vary in out-of-equilibrium driving
~\cite{calabrese2005evolution, calabrese2007quantum, hartman2013time, liu2014entanglement1, liu2014entanglement2, alba2017entanglement, 
mezei2017entanglement, tonni2018entanglement, wen2018entanglement, von2018operator, gong2018topological, rakovszky2019sub, alba2019quantum, gong2019lieb}. 
If the thermalization is triggered by a global unitary quench with an interacting post-quench Hamiltonian, 
the volume-law entropy will replace the area-law behavior~\cite{calabrese2005evolution, hartman2013time, mezei2017entanglement}.
Many factors may affect the non-equilibrium dynamics;
for instance, randomness may lead to the many-body localization (MBL) 
~\cite{basko2006metal, vznidarivc2008many, bardarson2012unbounded, iyer2013many, kim2013ballistic, vosk2013many, serbyn2013universal, 
huse2014phenomenology, serbyn2014quantum, vosk2015theory, singh2016signatures, khemani2017two, luschen2017observation, bordia2017probing, nahum2018dynamics}.
As a result of avoiding thermalization in the MBL, 
the stationary state (quench steady state) exhibits area-law entropy for the short-entangled systems, 
and the entanglement entropy grows logarithmically in time, 
which is in contrast with the linear growth in thermalized case.

Except for the disorder, 
the presence of non-unitary operations (e.g. relaxation, dissipation, measurements) 
will also influence the entanglement dramatically.
One notable example is, by introducing a continuous monitoring (damping) term in quench dynamics of free Fermion chain, 
it is found that the volume law entanglement is unstable 
under arbitrary weak damping if it is applied everywhere~\cite{cao2018entanglement}.
One related problem is the quantum Zeno effect~\cite{degasperis1974does, misra1977zeno}, where the total system is measured continuously and localized near a trivial product state.
Instead of above continuous monitoring, another protocol is to apply the local measurements discretely into quantum dynamics, with a finite probability $P$ named by \emph{measurement rate}.
One may expect that the presence of the local measurements will destroy the volume law produced by pure unitary dynamics, especially in low-dimensional systems.

However, this is not the case in generic interacting systems, where more subtle entanglement structures could survive under measurements.
A stable volume law phase was found under finite small measurement rate in several numerics on the circuit models~\cite{skinner2018measurement, li2018quantum, li2019measurement, choi2019quantum}.
By simulating hundreds of qubits in Clifford circuits, 
it is found a continuous quantum dynamical phase transition by tuning the measurement rate
~\cite{li2018quantum, li2019measurement}.
The presence of the volume-to-area law transition is also identified in Floquet and random unitary circuits \cite{skinner2018measurement},
and can be understood by a classical percolation problem.
Furthermore, it has been argued the presence of the volume-to-area law transition can be reinterpreted in a quantum error correction point of view \cite{choi2019quantum}.
The authors of Ref. \cite{choi2019quantum} considered the influence of two parameters -- depth and fraction -- 
on entanglement dynamics rather than considering only the measurement rate (density of non-unitaries in whole dynamics, which is equivalent to the fraction when the depth is one).
By tuning the additional parameter depth, they show that information scrambling plays a crucial role in the transition.
One can also directly define the strength of measurement, e.g., by introducing positive operator-valued measurements
~\cite{szyniszewski2019entanglement}.
Interestingly, the transition occurs at a finite measurement strength even for permanent measurements ($P=1$) \cite{szyniszewski2019entanglement}.

It should be noticed that this is not the end of the story.
Although the volume-to-area law transition is observed in several different circuit models numerically,
it is still unclear under what condition the volume law phase could survive.
An important example of absence of the transition is the non-interacting Bell pair model reported in Ref.~\cite{chan2018weak},
where arbitrary measurement rate exhibits an area law entanglement.
They also showed that by extending the Bell pair with only two-body entanglement to mutually entangled clusters,
a volume-to-area law entanglement transition can exist.

Taken the above facts, another question immediately arises:
What is the nature of this measurement-induced entanglement phase transition?
In Ref.~\cite{skinner2018measurement}, 
by mapping the calculation of zeroth R\'enyi entropy to a classical percolation problem, 
a toy model describing the disentanglement process for unitary dynamics with measurements was provided.
The scale invariance in critical percolation system exhibits a logarithmic growth of entanglement,
and also leads to power-law decay correlations.
These behaviors are observed in the circuit models by numerics~\cite{skinner2018measurement, li2019measurement}.
Under the symmetry between time-like and space-like directions, 
the von Neumann entropy (the first R\'enyi entropy) is found to grow logarithmically in both space and time.
The mutual information is investigated as a measure of quantum correlations between two separated sites,
and exhibits power-law decay in space.
Based on these critical entanglement structures, an underlying conformal field theory (CFT) description was proposed. 
Moreover, it was shown that the peak of mutual information~\cite{li2019measurement} and 
the variance of entanglement entropy~\cite{szyniszewski2019entanglement} can be the indicator of the volume-to-area law phase transition,
which gives more insights on the possible statistic description of this transition.
Please note that, the above discussions are mainly based on the circuits model \cite{skinner2018measurement,li2019measurement}, 
and non-interacting models \cite{li2019measurement}. 
Besides, much less is known about
the universality (if any) of the entanglement entropy of more generic models, 
e.g., quantum many-body lattice systems, or of models mappable to them. 

In this paper, we study the quantum dynamics of one-dimensional Bose-Hubbard model in the presence of random projective measurements 
by using matrix product state (MPS) and time-evolving block decimation (TEBD)~\cite{suzuki1976generalized, vidal2004efficient, white2004real, daley2004time, paeckel2019time}. 
The Bose-Hubbard model has been a paradigmatic non-integrable lattice model 
to understand the quantum dynamics and non-equilibrium properties. 
We map out a global phase diagram controlled by the measurement rate in time domain $N_t$ and spatial space $P_x$ (see Fig. \ref{fig:network} for definition).
A volume-to-area law phase transition is observed,
and the region of volume law phase becomes wider with decreasing $N_t$.
The behavior provides strong evidence of the existence of a stable volume law phase, and also shows the importance of the information scrambling in this dynamical transition.
We find that the single-site entanglement entropy can indicate the transition,
similar to the case of disorder-induced MBL~\cite{wahl2019signatures}.
At the critical point, 
we obtain a logarithmic growth of entanglement and power-law decay of correlations. 
The scaling behavior of entropy around the critical point  appears to belong to a single universality class.
Our work provides a wealth of evidences that non-unitary factors, such as projective measurements,
can induce a dynamical phase transition, adding more pieces of message 
to the recently proposed theoretical scenario \cite{skinner2018measurement,li2019measurement,choi2019quantum},
from the microscopic view on non-integrable quantum lattice model.

\section{Model and Method}

\begin{figure}\centering
\includegraphics[width=\columnwidth]{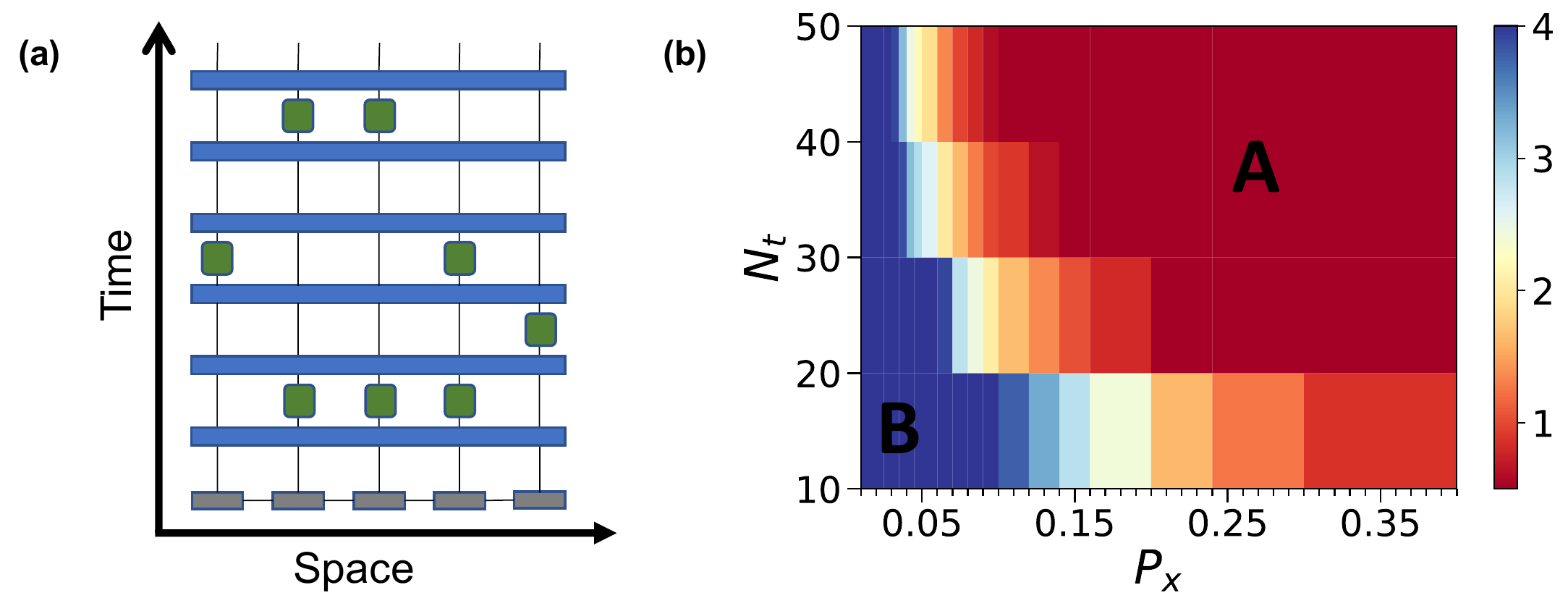}
\caption{ \label{fig:network} (a) A diagrammatic representation of the quench dynamics with local projective measurements. 
The quench dynamics is from a trivial product state described by MPS (gray rectangles), and the unitary time evolution background is build with several unitary layers described by MPO (blue rectangles). 
The local projective measurements (green rounded squares) are applied only after several unitary layers, which are chosen randomly with measurement rate in time $N_t$. For each measured layers, the local projective measurements are posited randomly with measurement rate in space $P_x$.
(b) Two-dimensional phase diagram of the entanglement phase transition as a function of the measurement rate in time $N_t$ and in space $P_x$, where $\bf{A}$ and $\bf{B}$ represent the disentangling and entangling phases respectively.
The background colors represent the value of the von Neumann entropy for long-time steady states with sub-system size $L = 8$ averaged over different (up to 900) random realizations. }
\end{figure}

In the present work, we consider a quench dynamics of the von Neumann entropy on an one-dimensional Bose-Hubbard model~\cite{fisher1989boson} as the post-quench Hamiltonian
\begin{equation}
H = - J_0 \sum_i (b^\dagger_i b_{i+1} + b^\dagger_{i+1} b_i) + \frac{U}{2} \sum_i n_i(n_i - 1)
\end{equation}
where $b^\dagger_i$, $b_i$, and $n_i = b^\dagger_i b_i$ are the boson creation, annihilation, and particle number operators on site $i$ respectively.
The model has a critical point $U_c / J_0 \sim 3.3$ \cite{kashurnikov1996exact, kashurnikov1996mott, kuhner2000one, zakrzewski2008accurate, lauchli2008spreading, ejima2011dynamic, pino2012reentrance, ejima2012characterization, rachel2012detecting, carrasquilla2013scaling}, 
which separates a superfluid phase in $U < U_c$ from a Mott insulator phase in $U > U_c$.
Our setup for the unitary time evolution background is a quench dynamics from the Mott phase into the superfluid phase.
Without losing the generality, in our simulation we set $J_0 = 1$ and $U = 0.14$,
and the initial state is chosen to be a trivial product state with occupation number on each site $n_i = 1$.
The maximum number of bosons per site is set to be $5$.

The quench dynamics is simulated using MPS and TEBD performing in TeNPy package \cite{hauschild2018efficient}, and a network diagrammatic representation is shown in Fig.~\ref{fig:network}(a).
The unitary time evolution background is built by several layers of the matrix product operator (MPO), 
describing the time evolution operator. 
By performing TEBD, each unitary layer is written in terms of two-site gates by using a second order Suzuki-Trotter decomposition with $dt=0.02$.
A bond dimension $\chi$ up to 2048 was test to be fine for the time scale (up to $T = 30$) considered in the present work. 
In our calculation, an open boundary chain with total system size $L_0 = 36$ is considered. 
We also test some calculations on $L_0=48$ size and get the very similar behavior, 
which gives us confidence that the results shown below are free of the finite-size effect.

The local projection operator is defined by $\mathcal{O}_P=|1_x\rangle \langle 1_x|$,
which projects the quantum state at spatial position $x$ to $|1\rangle$.
These operators are set to be applied randomly in the dynamics.
We introduce the measurement rate in time domain $N_t$ and spatial space $P_x$ separately,
as the tuning parameters in our model.
There are $N_t$ measurement layers per time unit (we set time unit $\hbar/J_0=1$) in the time evolution process.
In our setup, for each time unit we generate a list with length $1/dt = 50$.
For the considered time scale $T = 30$, there are 30 such lists,
which build one-to-one correspondence with the unitary layers in the network.
Each list contains $50 - N_t$ ``0'' and $N_t$ ``1'' values,
and the order of these ``0'' and ``1'' values is set to be random.
If the random number is ``1'', the local projective measurements will be applied after this layer; otherwise no measurement is applied.
Here, we note that the lists for each time unit are generated independently,
so that the measured layers are selected randomly, not in a periodical pattern.
For each measured layer, we define the probability applying local projective measurements on a single site to be the measurement rate in spatial space $P_x$.
Numerically this is achieved by generating a list of random numbers $R_x$ in uniform distribution between 0 and 1.
If $R_x < P_x$, the spatial position $x$ will be measured; otherwise it is unmeasured.
After each local measurement, the total state has been renormalized, and therefore the full dynamics is nonlinear. 
For smaller measurement rates, we have simulated more random realizations;
for example, for $P_x = 0.02$, $N_t = 50$ we simulate 900 random realizations, 
since the effect of single local projective measurement is larger. 
For larger measurement rates, the effect of single local projective measurement is much smaller, 
so we do not need many random realizations to obtain the smooth averaged curve with small standard error; for example, for $P_x = 0.08$, $N_t = 50$ we only simulate 40 random realizations.

The two parameters $N_t$ and $P_x$ considered in our simulations are independent.
By definition, the average number of measurement per site per unitary time $N_{\rm aver} = N_t \ P_x$.
In fact, the $N_t$ considered in our simulation controls the degree of information spreading effectively, 
similar to the ``depth'' defined in Ref.~\cite{choi2019quantum} but with randomness.

\section{Results and discussions}

\begin{figure}\centering
	\includegraphics[width=\columnwidth]{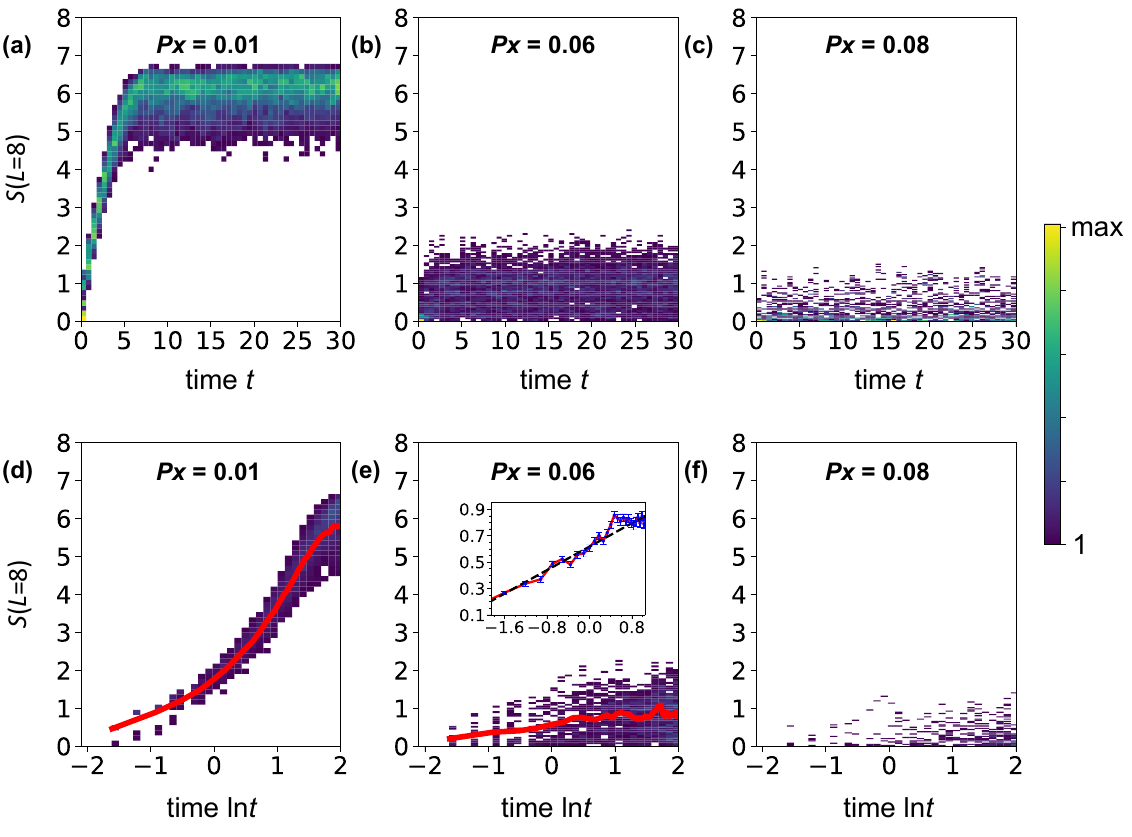}
	\caption{\label{fig:logt} 
		The dynamics of the von Neumann entropy with sub-system size $L = 8$ for different values of $P_x$;
		here the results for $N_t=50$ are chosen.  The results for different (up to 900) random realizations are presented in the form of a two-dimensional color-coded histogram. 
		(a-c) The full dynamics of the entropy with time up to $T = 30$, where the von Neumann entropy already reaches the platform saturated by the sub-system size. 
		(d-f) The short-time dynamics of the entropy plotted in $\ln t$. The red line shows a $\ln t$ growth of the averaged entropy. The inset in panel (e) shows a fitting in form $S=a \ln t + b$, with $a \approx 0.22 \pm 0.02$ and $b \approx 0.58 \pm 0.02$. } 
\end{figure}

By examining the von Neumann entropy, we observe the competing tendencies between the local projective measurements and 
unitary time evolution background, which suggest a phase transition between entangling and disentangling phases.
A two-dimensional phase diagram of von Neumann entropy consisting with the measurement rate 
in time $N_t$ and space $P_x$ is presented in Fig.~\ref{fig:network}(b).
In the case of $N_t = 50$, the region of volume law phase in our model is very narrow.
Because of this, one may question about stability of the strong entangling phase in a combination of unitary evolution and local projective measurements.
As shown in Fig.~\ref{fig:network}(b), for different considered values of $N_t = 50, 40, 30, 20$, the corresponding critical spatial measurement rate $P_{x,c} \approx 0.06, 0.08, 0.12, 0.24$ respectively.
The fast (than linear) growth of $P_{x,c}$ with decreasing $N_t$ strongly supports the existence of a phase transition when $N_t$ is finite small.
One may naively think that, the phase diagram is only controlled by the density of the local projective measurements applied $N_{\rm aver} = N_t \ P_x$.
In our setup, we show that the degree of information scrambling (controlled by $N_t$) also plays an important role and contributes to the critical value of $N_{\rm aver}$.
Our numerical results for different $N_t$ show that, with deceasing $N_t$, the critical value of $N_{\rm aver}$ increases.
This, in fact, is the consequence of the competing tendency between density of the local projective measurements applied and the degree of information scrambling.
Obviously, the critical value of $N_{\rm aver}$ is model dependent, since the information scrambling will be changed in different model realizations.
We also note that, although the critical $N_{\rm aver}$ is not universal, the behaviors at or near critical point is expected to be universal in different models.
In further discussions, we will focus only on the case of $N_t = 50$.

The difference between the entanglement structures in two phases is evident in Fig. \ref{fig:logt}(a-c),
where we show the time evolution of the entropy for various values of $P_x$.
For the smaller measurement rate (Fig. \ref{fig:logt}(a)),
the entanglement entropy increases ballistically at initial times and then saturates at a large value.
The entropy in the quasi-stationary regime at long times is close to the volume-law values in the unitary evolution.
For larger measurement rate (Fig. \ref{fig:logt}(c)), the entropy saturates quickly to very small values,
corresponding to an area law phase or localized phase.

\begin{figure}\centering
	\includegraphics[width=\columnwidth]{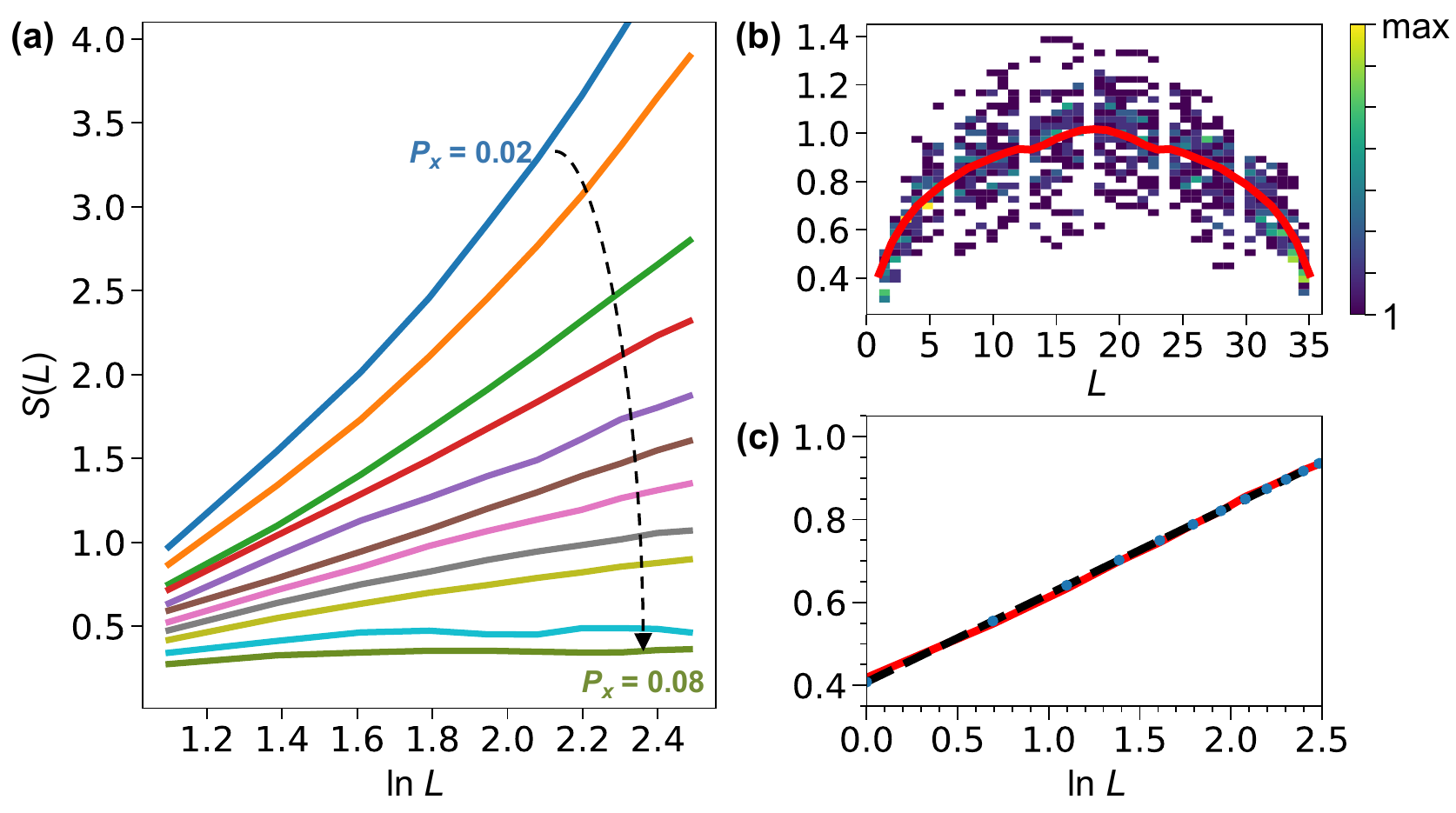}
	\caption{\label{fig:logL} (a) The spacial distribution of the von Neumann entropy for long-time steady states, with fixed $N_t = 50$, 
		after averaging over hundreds of random realizations. Different curves from top to bottom correspond to different values of 
		$P_x = 0.02, 0.025, 0.03, 0.035, 0.04, 0.045, 0.05, 0.055, 0.06, 0.07, 0.08$.
		(b) The von Neumann entropy of long-time steady states for $P_x \sim P_{x,c}$. 
		The results for 240 different random realizations are presented in the form of a two-dimensional color-coded histogram. The red line shows the averaged data.
		(c) The red line shows the averaged von Neumann entropy of long-time steady states for $P_x \sim P_{x,c}$ plotted in $\ln L$. The black line shows the result of a fitting in the form $S=a \ln L + b$, with $a \approx 0.21 \pm 0.01$ and $b \approx 0.41 \pm 0.01$. } 
\end{figure}

Besides the long-time behaviors exhibit a phase transition between volume law and area law phases, 
the short-time behaviors (Fig.~\ref{fig:logt}(d-f)) also display interesting features. 
In the area law phase (Fig. \ref{fig:logt}(f)), 
the state is close to a trivial product state due to the very frequent measurement,
therefore the entropy is very small and almost unchanged in time. 
In the volume law phase (Fig. \ref{fig:logt}(d)), 
the entanglement growth exhibits a linear dependence (with possible logarithmic correlations) with time like in the case of pure unitary evolution, 
and the growth velocity decreases with increasing the measurement rate.
At the critical point (Fig. \ref{fig:logt}(e)), the linear growth is totally destroyed by the local measurements, 
the entanglement exhibits a logarithmic growth in time as shown in Ref.~\cite{skinner2018measurement} for zeroth R\'enyi entropy.

Under the symmetry between space-like and time-like directions~\cite{skinner2018measurement},
we also expect to observe the phase transition by the entanglement growth in space (sub-system size).
In Fig.~\ref{fig:logL}(a), the averaged von Neumann entropy of long-time steady states for different $P_x$ is presented.
The growth velocity of the von Neumann entropy in space decreases as a result of increasing measurement rate. As shown in Fig.~\ref{fig:logL}(b), at the critical point, a logarithmic dependence of the entropy on the sub-system size is observed.
Our numerical results of the spacial distribution of the entropy suggests the scaling behavior introduced in Refs.~\cite{vasseur2018entanglement, skinner2018measurement, li2019measurement}, where a logarithmic correction is added onto the linear dependence in the volume law phase.
The logarithmic growth entropy emerges at the criticality both in space and time, 
in line with the percolation picture proposed in Ref.~\cite{skinner2018measurement}, where the logarithmic scaling is proved~\cite{chayes1986critical}.
As a signature of scale invariance, the logarithmic growth entropy also implies a possible underlying CFT description~\cite{vasseur2018entanglement, skinner2018measurement, li2019measurement}.
Moreover, in the critical percolation problem, another important signature of scale invariance is the the power law decay correlations, which are also observed in our model (see below).

\begin{figure}
	\includegraphics[width=\columnwidth]{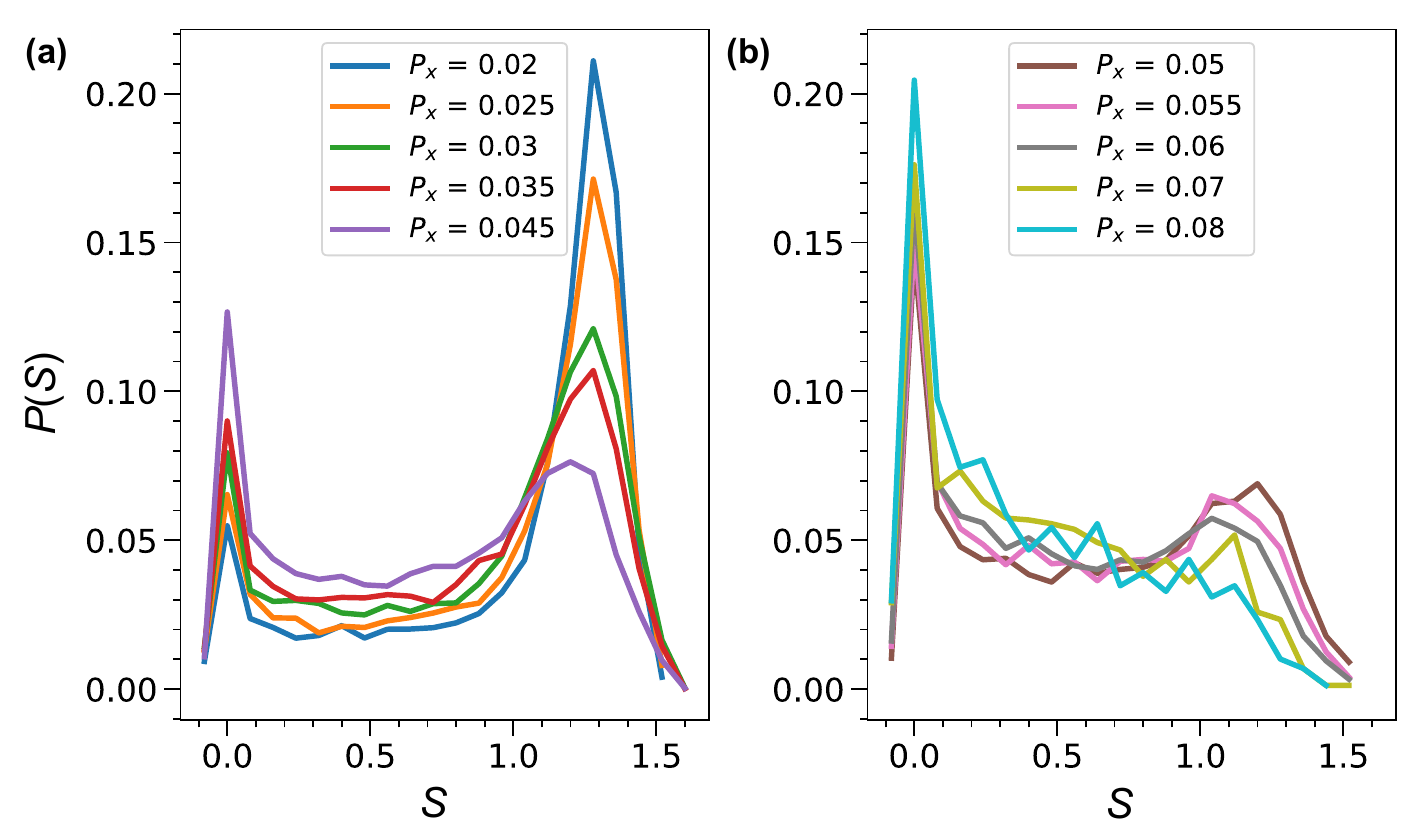}
	\caption{\label{fig:SSEE} Probability distribution of the single-site von Neumann entropy for different values $P_x$, with fixed $N_t = 50$. 
		(a) Deep in volume-law phase and (b) Near the critical point and in area-law phase. }
\end{figure}

Before discussion on the critical scaling, we introduce another indicator of the measurement-induced volume-to-area law transition, called \emph{single-site von Neumann entropy}.
Its probability distribution for different $P_x$ is plotted in Fig.~\ref{fig:SSEE}.
We find that the single-site entanglement entropy raises signature of the transition similar to the case of the thermal-to-MBL transition~\cite{wahl2019signatures}. 
In the volume law phase, as shown in Fig.~\ref{fig:SSEE}(a), the single-site entropy distribution hosts two dominate peaks. 
One is located at $S=0$ and the other one is at $S\approx 1.25$.
When we increase the measurement rate, the peak centered at $S>0$ becomes broadened, and the $S=0$ peak becomes sharper.
For the case of the measurement rate near the critical value, as the curves for $P_x = 0.05$ and $0.055$ shown in Fig.~\ref{fig:SSEE}(b), 
the $S>0$ peak is very small. In the area law phase, the $S>0$ peak totally disappears. 
That a single quantity can indicate both measurement-induced volume-to-area law transition and disorder-induced thermal-to-MBL transition, 
suggests a possible generic picture of entanglement transition.
Considering a pure unitary dynamics without any impurity effect (e.g. disorders and measurements),
the local information spreads into the whole space, 
so that the single-site entanglement is strong and only one peak should appear in $S>0$ side.
The presence of disorders/measurements generally stabilizes quantum information in the system, resulting in a peak near vanishing entropy.

\begin{figure}\centering
	\includegraphics[width=\columnwidth]{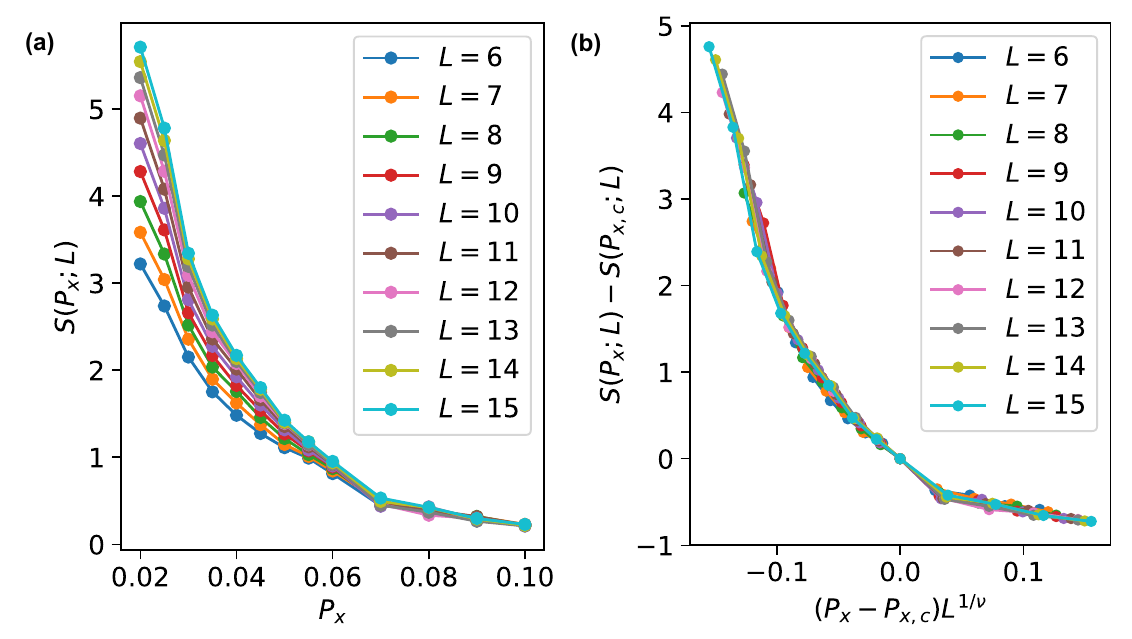}
	\caption{\label{fig:scaling} (a) The von Neumann entropy of long-time steady states with different sub-system size $L$ as a function of the measurement rate in space $P_x$, with fixed $N_t = 50$. (b) Data collapse of the same data presented in panel (a) by scaling form in Eq.~\ref{eq:scaling} with $P_{x,c} = 0.06$ and $\nu=2$.} 
\end{figure}

We now turn to consider global scaling behavior of entanglement entropy.
In order to extract critical behavior around the critical measurement rate $P_{x,c}$, 
we perform a finite-size scaling form for the von Neumann entropy~\cite{vasseur2018entanglement, skinner2018measurement, li2019measurement, choi2019quantum}
\begin{equation}\label{eq:scaling}
S(P_x) - S(P_{x,c}) = F((P_x - P_{x,c}) L^{1/\nu})
\end{equation}
where $L$ is the sub-system size. The data collapse yields the critical value $P_{x,c}\approx0.060 \pm 0.004$ and the exponent $\nu \approx 2.00 \pm 0.15$.
The fine data collapse presented in Fig.~\ref{fig:scaling}(b) shows the correctness of the scaling form, 
and also strongly supports the universal phase transition.
When $P_x$ asymptotically reduces to $P_{x,c}$, there is a region of logarithmic scaling entropy (with fluctuations), as shown in Fig.~\ref{fig:logL}.
This subtle structure is suddenly canceled out when $P_x > P_{x,c}$, which also indicates a critical point.
In particular, the obtained scaling index $\nu \approx 2.00$ in our calculation is close to the dynamics of random unitaries in one dimension \cite{skinner2018measurement}. 
The emergence of such a consistency implies that the randomness of local measurements and unitary gates may have similar origins.
Moreover, percolation criticality yields $\nu = 4/3$ in the bulk theory, which is different from our simulation and the result in random unitary circuits.
Interestingly, Ref.~\cite{vasseur2018entanglement} argues that the physical $\nu$ is defined by boundary theory,
where the Knizhnik-Polyakov-Zamolodchikov formula~\cite{knizhnik1988fractal} gives $\nu = 2$. 
In this regard, our numeric estimation seems matches the percolation picture quite well.

\begin{figure}
	\includegraphics[width=\columnwidth]{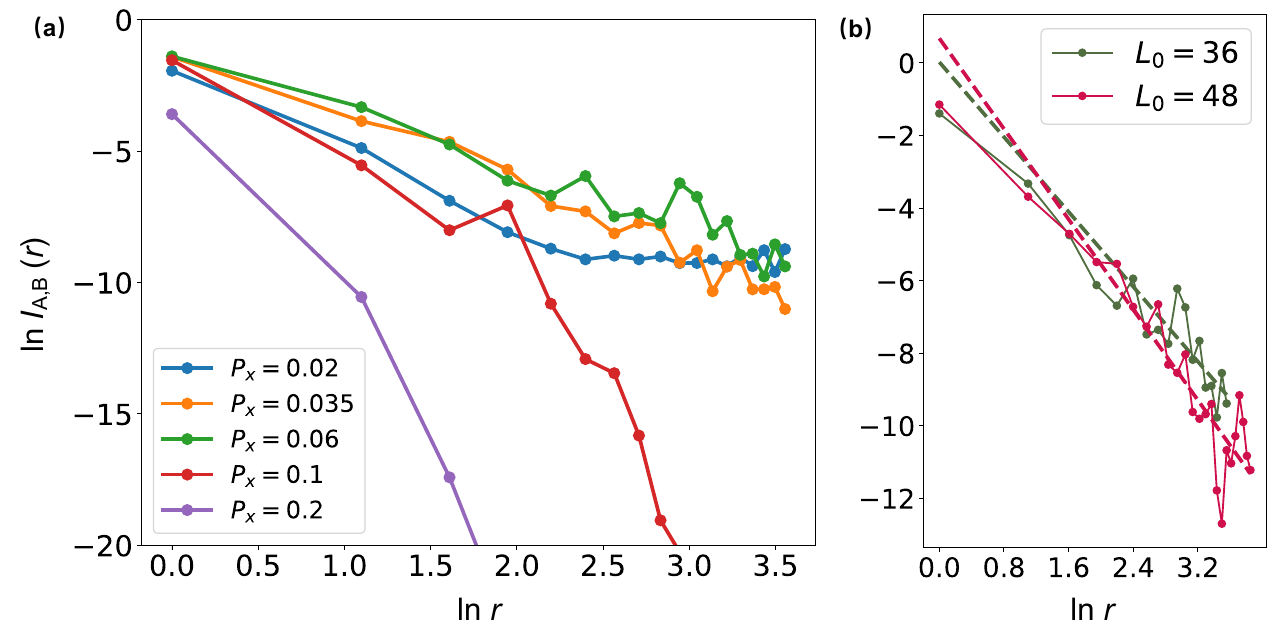}
	\caption{\label{fig:mutinf} Decay of the the mutual information $I_{A, B}$ in spatial distance $r$. 
		(a) The mutual information $I_{A, B}$ as a function of $\ln r$ for different values of $P_x$, with fixed $N_t=50$. 
		(b) Linear fitting of the mutual information in form $\ln I_{A, B} = a\ln r + b$ with the critical measurement rate $P_x = 0.06$. The data for different total system sizes $L_0 = 36$, $48$ are plotted by green and red lines, respectively. The fitting results are $a(L_0 = 36) \approx -2.58 \pm 0.65 $, $b(L_0 = 36) \approx 0.02 \pm 1.99$ and $a(L_0 = 48) \approx -3.12 \pm 0.57$, $b(L_0 = 48) \approx 0.68 \pm 1.89$, which give critical exponents $\Delta(L_0 = 36) \approx 1.29$ and $\Delta(L_0 = 48) \approx 1.56$. } 
\end{figure}

As we mentioned above, the power-law decay correlations in spatial space is served as a signature of the scale invariance.
Next we consider the correlation of entanglement by investigating the mutual information between two distant sites $A$ and $B$
\begin{equation}
I_{A, B} = S_A + S_B - S_{A \cup B} \ .
\end{equation}
From the enhanced conformal invariance, one expects that the critical exponent of power-law decay correlation is $\Delta = 2$ \cite{skinner2018measurement, li2019measurement}, 
which leads to $I_{A, B} \propto r^{-4}$. A comparison between different values of measurement rate $P_x$ is presented in Fig.~\ref{fig:mutinf}(a).
Note first that, the mutual information decays slowest at the critical point.
When $P_x$ is large (after transition), the measurements applied very frequently,
and the correlation decays exponentially since the long-time steady state is close to a product state. 
When $P_x$ is small (before transition), the unitary time evolution leads the system to thermalization, where the local information vanishes, so that results in a fast decay of correlations.
At the critical point, the mutual information between two sites $I_{A, B}$ with long distance is almost unchanged with increasing distance, 
and $I_{A, B}$ for critical value of $P_x$ exhibits a power-law decay.
As shown in Fig.~\ref{fig:mutinf}(b), the linear fitting in form $\ln I_{A,B} = a \ln r + b$ results in a critical exponent 
$\Delta = -a/2 \approx 1.29$ for total system size $L_0 = 36$. 
On larger system size ($L_0=48$), it is found this value enhances to $\Delta \approx 1.56$. 
As discussed above, the obtained critical exponent from CFT is $\Delta=2$ \cite{skinner2018measurement, li2019measurement}.
Through this comparison, our result seems non-universal,  
which can be attributed to the following reasons. 
First, unlike the circuits model, the current calculations are based on 
a non-integrable quantum lattice model.
Second, our calculations are limited to finite sizes, and the non-universal behavior is a finite-size effect.

\section{Summary and outlooks}

In this paper, we have studied the entanglement dynamics of a 1D Bose-Hubbard model with local projective measurements randomly appearing in space and time.
A volume-to-area law entanglement phase transition was observed. By considering two parameters -- measurement rate in time $N_t$ and space $P_x$, 
a two-dimensional phase diagram was presented. It is found that the volume law phase is robust with local measurements applied.
The time dependence and spatial distribution of the entanglement entropy distinguish the two phases,
and the single-site entanglement entropy is introduced as another indicator of this transition.
Finite-size scaling analysis indicates that the transition falls into a single universality class, 
and the critical exponent $\nu \approx 2$ 
is close to the estimation in random unitary circuits \cite{skinner2018measurement}.

For the critical values of measurement rate, a logarithmic growth of entanglement entropy on sub-system size and evolved time was observed.
Moreover, the mutual information between two distinct sites was found to exhibit a power-law decay in space. 
Both the logarithmic growth entanglement entropy and the power-law decay correlations support the presence of a scale invariant quench steady state. 
Based on this, our results support that it is possible to describe the observed volume-to-area law phase transition by a conformal field theory \cite{vasseur2018entanglement,skinner2018measurement,li2019measurement}.

The present work opens a number of questions on the entanglement phase transition.
For example, it is important to find more common signatures of different transitions as the single-site entanglement entropy investigated in our work.
These information can help to build a general picture of entanglement phase transition.
Moreover, in Ref.~\cite{vasseur2018entanglement}, to describe general entanglement transition, e.g. MBL transition,
a random tensor network (RTN) picture is proposed in a holographic way.
In fact, the volume-to-area law transition in RTN is closer to the measurement-induced phase transition investigated in the present work.
Importantly, the critical exponent $\nu \approx 2$ found in our work and previous numerics \cite{skinner2018measurement}
is very close to the theoretical value predicted from the boundary of percolation bulk theory~\cite{vasseur2018entanglement}.
More detailed study of a possible connection to the holographic picture is deserved to explore in the future.

For the measurement-induced transition, Refs.~\cite{bao2019theory, jian2019measurement}
provide theoretical understanding of the transition by mapping onto a percolation problem within a replica method.
Based on the fact that the conformal method can be used to describe critical two-dimensional percolation, they argue for an underlying CFT of the entanglement transition.
However, the exact mapping to the percolation model only works under the limit of large local Hilbert space dimension $\mathcal{D}$.
In more realistic cases with finite $\mathcal{D}$, the situation is still unclear.

It would be also interesting to investigate the role of randomness in entanglement transition.
In thermal-to-MBL transition, the randomness is not necessary; a quasi-periodic potential~\cite{iyer2013many} also can lead to failure of thermalization.
In the case of measurement-induced entanglement transition, the role of randomness is still unclear.
In Ref.~\cite{li2019measurement}, it was found that the randomness of the unitary gates and the local measurements is inessential by simulating Clifford circuits. 
But the numerics of Harr circuits in Ref.\cite{skinner2018measurement} show that the randomness of the unitary gates leads to a different critical exponent, 
which means a different universality. 
We leave the discussion of the possible non-randomness induced entanglement transition for future work.

\begin{acknowledgements}
W.Z. thanks X. Chen and Q. T. thanks Q.-H. Guo for discussion.
This work is supported by start-up funding from Westlake University. 
The numerical calculations in this paper have been done on the supercomputing system in the Information Technology Center of Westlake University.
\end{acknowledgements}

\bibliographystyle{apsrev4-1}
\bibliography{apssamp}

\end{document}